\documentclass{article}
\usepackage[utf8]{inputenc}
\usepackage{spconf,amsmath,graphicx}
\usepackage{url}

\DeclareMathOperator*{\argmin}{arg\,min}

\usepackage{bm} 

\newcommand{\figref}[1]{Fig.~\ref{#1}}

\newcommand\vb{\ensuremath{\mathbf{b}}}
\newcommand\vc{\ensuremath{\mathbf{c}}}

\newcommand\vs{\ensuremath{\mathbf{s}}}

\newcommand\vy{\ensuremath{\mathbf{y}}}

\newcommand\valpha{\ensuremath{\bm{\alpha}}}

\newcommand\mW{\ensuremath{\mathbf{W}}}

\newcommand\mTheta{\ensuremath{\bm{\Theta}}}
\newcommand\mPsi{\ensuremath{\bm{\Psi}}}

\title{Demodulating Subsampled Direct Sequence Spread Spectrum Signals Using Compressive Signal Processing}
\name{Karsten Fyhn, Thomas Arildsen, Torben Larsen and Søren Holdt Jensen
\thanks{First published in the Proceedings of the 20th European Signal Processing Conference (EUSIPCO-2012) in 2012,
published by EURASIP. This work is supported by The Danish Council for Strategic Research under grant number 09-067056.}}
\address{Department of Electronic Systems, Aalborg University, Denmark. \\\{kfn,tha,tl,shj\}@es.aau.dk}

\begin{document}
\maketitle

\begin{abstract}
We show that to lower the sampling rate in a spread spectrum communication system using Direct Sequence Spread Spectrum (DSSS), 
compressive signal processing can be applied to demodulate the received signal.
This may lead to a decrease in the power consumption or the manufacturing price of wireless receivers using spread spectrum technology.
The main novelty of this paper is the discovery that in spread spectrum systems it is possible to apply compressive sensing with a much simpler hardware architecture than in other systems,
making the implementation both simpler and more energy efficient.
Our theoretical work is exemplified with a numerical experiment using the IEEE 802.15.4 standard's $2.4\:$GHz band specification.
The numerical results support our theoretical findings and indicate that compressive sensing may be used successfully in spread spectrum communication systems.
The results obtained here may also be applicable in other spread spectrum technologies, such as Code Division Multiple Access (CDMA) systems.
\end{abstract}


\section{Introduction}
The concept of compressive sensing \cite{Candes2006c,Donoho2006} is attracting more and more attention in the signal processing community.
Where the classical Shannon-Nyquist sampling theorem requires a signal to be sampled at twice its signal bandwidth,
compressive sensing samples the signal at its information rate, which may be much lower.
Compressive sensing is used to reconstruct a signal to a full Nyquist rate representation, but if only inference about information in the signal is desired,
compressive signal processing is better suited \cite{Davenport2010}.
Compressive signal processing is used when inference about information in a signal is of interest, rather than the reconstruction of the signal itself.
Compressive sensing and compressive signal processing samples the signal using a sampling scheme with typically a randomized structure and then exploits sparsity in the signal to enable subsampling.
In DSSS systems the sparsity is in the selection of a code used for transmission of a given data sequence.
In this work we show how compressive signal processing may be applied to a spread spectrum receiver to lower the sampling rate at the receiver.
This may lower the overall energy consumption of the device and/or lower the price of the Analog to Digital Converter (ADC).
To exemplify this consider the following: This work is based on a signal model used in the IEEE 802.15.4 standard \cite{IEEE802.15.4}, in which a baseband signal with a Nyquist frequency of $200$kHz must be sampled.
To show the benefit of lowering the sampling rate, we compare two ADCs from Analog Devices\footnote{\url{http://www.analog.com}}: The AD7819 and the AD7813.
The AD7819 is an 8-bit ADC with a maximum throughput of 200 kilosamples per second, whereas the AD7813 is an 8- or 10-bit ADC with a maximum throughput of 400 kilosamples per second.
We are aware that 400 kilosamples per second is the Nyquist rate of the system and the sampling rate should be higher than this to comply with the Shannon-Nyquist sampling theorem.
However, we use these two ADCs as they are almost identical in every aspect except for the sampling rate, making them perfect for comparison.
In present IEEE 802.15.4 compliant receivers, an ADC similar to the AD7813 must be used to comply with Shannon-Nyquist, 
but if compressive signal processing is able to lower the sampling rate by a factor of two,
the AD7819 may be used instead. 
These two particular ADCs use the same amount of power so there are no energy savings, but where the AD7813 costs $2.98$\$,
the AD7819 only costs $2.29$\$.

Previous work has studied the use of compressive sensing in Ultra-Wideband (UWB) systems for channel estimation where the sparsity of the signal lies in the time domain \cite{Paredes2007a,Zang2009}.
Some researchers have studied the use of compressive sensing for spread spectrum communication systems \cite{Aggarwal2009}.
However, this work is mainly theoretical and relies on second order Reed-Muller codes, which would be difficult to implement in hardware.
A more practical approach is given in \cite{Li2011a} where compressive sensing is used to decrease the sampling rate of a GPS receiver by exploiting sparsity in the number of possible signal components at the receiver.
However, this approach also suffers from a complicated hardware implementation.
In both works the receiver must use complicated hardware filters, which may make their implementation very difficult, 
considering the impact of hardware filters on compressive sensing performance \cite{Pankiewicz2011}.
In this work we apply compressive signal processing to a general DSSS system.
We show that in a spread spectrum system it is possible to use simply a repeated version of the matched filter used in classic receivers
instead of using a complicated filter structure to acquire random measurements.
This greatly simplifies the implementation and makes compressive sensing feasible for implementation in spread spectrum wireless receiver systems.
Our approach is not limited to DSSS but may also be applied in other spread spectrum technologies, such as CDMA.

One major obstacle in applying compressive sensing to any wireless system is the presence of noise folding,
which occurs because the noise is not measurement noise, but noise added before measuring the signal.
This severely impacts the receiver performance, which is also evident in our numerical experiments.
We discuss how to mitigate this in Section \ref{sec:discAndConc}.

In the following we first define classic transmitter and receiver structures in Sections \ref{sec:txStruc} and \ref{sec:rxStruc}, respectively.
Then we show how the classic receiver structure must be modified to incorporate compressive signal processing in Section \ref{sec:csStruc}.
Our theoretical work is exemplified with a numerical experiment using the IEEE 802.15.4 standard in Section \ref{sec:numExp} followed by a discussion and conclusion in Section \ref{sec:discAndConc}.

\section{Transmitter Structure}
\label{sec:txStruc}
In both the transmitter and the receiver structure we treat the signal symbol-by-symbol,
where each symbol may be a single bit of information or a block of bits.
Let $\vb_k \in \{\pm 1\}^{\mathrm{N\times 1}}$ be a binary vector, signifying the $k$th symbol to be transmitted and consisting of $N$ information bits.
Now define a binary pseudo-random noise (PRN) sequence as $\vc_k \in \{\pm 1\}^{\mathrm{C\times 1}}$.
These two binary vectors are the discrete equivalents of an information signal and a PRN signal, $b_k(t)$ and $c_k(t)$, respectively as shown in \figref{fig:transceiverStruc} and are defined as:
\begin{figure*}
	\centering
	\includegraphics[width=0.80\textwidth]{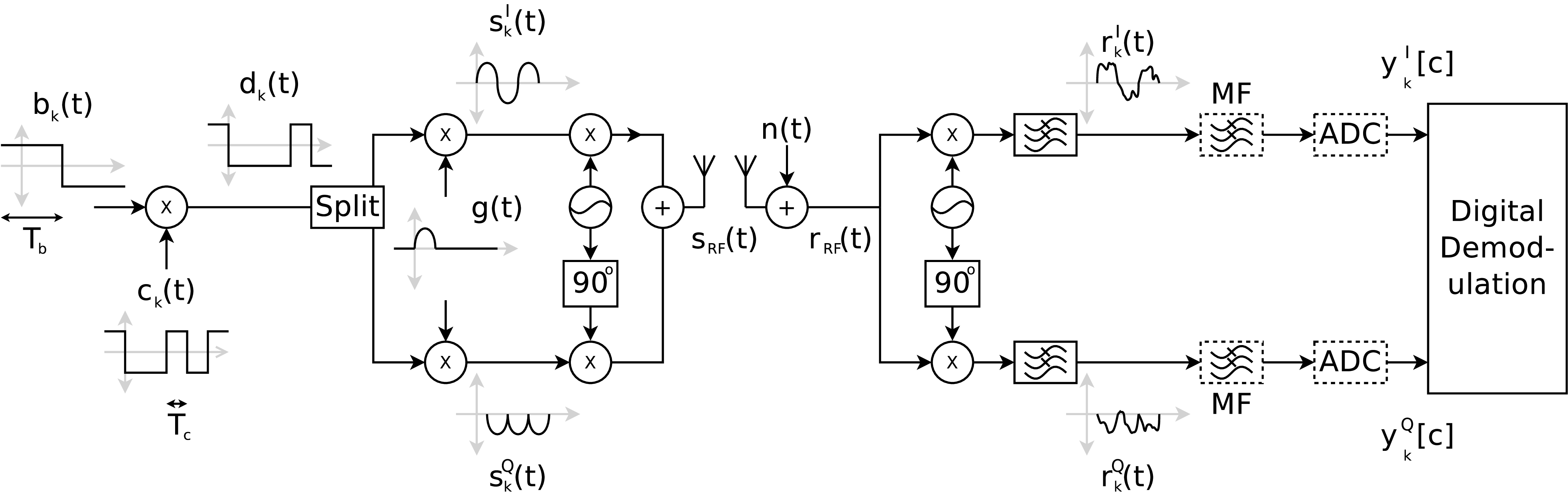}
	\caption{Transmitter and receiver structure for QPSK modulation/demodulation. The items drawn using dotted lines are hardware components that must be modified to enable compressive sensing.}
	\label{fig:transceiverStruc}
\end{figure*}
\begin{align}
	b_k(t) &= \sum_{n=0}^{N-1} \vb_k[n]\mbox{rect}\!\left(\frac{t-nT_b}{T_b}\right),~0 \leq t < NT_b,\\
	c_k(t) &= \sum_{c=0}^{C-1} \vc_k[c]\mbox{rect}\!\left(\frac{t-cT_c}{T_c}\right),~0 \leq t < CT_c,
\end{align}
where $T_b$ and $T_c$ are the bit and chip duration, respectively, and $NT_b=CT_c$.
We define:
\begin{align}
    \mbox{rect}\!\left(t\right)=\Big\{\begin{matrix} ~1 & \text{if } 0 \leq t < 1,\\ ~0 & \text{otherwise.}~~~~\end{matrix}
\end{align}
When multiplied, they form the spread spectrum data signal, $d_k(t) = b_k(t)c_k(t),~0\leq t < NT_b$.

The notation used in the above may in some cases be simplified, as the choice of a PRN sequence might be implemented as a mapping from one bit or a block of bits directly to a given sequence of chips,
as done in e.g. IEEE 802.15.4 \cite{IEEE802.15.4}.
In the following, the signal model we define is based on the IEEE 802.15.4 standard's $2.4~$GHz band specification.
This means the encoding using DSSS may be written as a matrix-vector product, with $M=2^N$ possible data signals:
\begin{align}
	d_k(t) &= \mPsi(t) \valpha_k, \text{ where}\\
    \mPsi(t) &= \begin{bmatrix} d_1(t) \\ d_2(t) \\ \vdots \\ d_{M}(t) \end{bmatrix}^\mathrm{T}\!\!\!\!,~0\leq t <NT_b,
\end{align}
where $\mPsi(t)$ is a dictionary of possible data signals and $\valpha_k \in \{0,1\}^{\mathrm{M}\times 1}$ is a sparse vector with only one non-zero entry, 
namely the entry that selects a given PRN sequence from the dictionary.
It may also be considered a symbol vector as it corresponds to the k$th$ symbol being transmitted. 
The sparsity of $\valpha_k$ is what enables us to use compressive sensing for demodulation.
The sparsity of the signal lies in which PRN sequence is chosen for transmission.

The IEEE 802.15.4 $2.4~$GHz band specification is based on QPSK and therefore the output sequence is split up, so that even-indexed chips in $d_k(t)$ are transmitted in the in-phase path and odd-indexed chips in the quadrature-phase path.
In the following we only state the equations for the in-phase path, but similar expressions may be derived for the quadrature-phase part.
The resulting data signals are then used to modulate some pulse shape function, $g(t)$:
\begin{align}
	s_{k}^I(t) &= \mPsi^I(t) \valpha_k,\text{ where}\\
	\label{eqn:IDic}\mPsi^I(t) &= \begin{bmatrix} \sum\limits_{c \in S} d_1(t)g\left(t-cT_c\right) \\ \sum\limits_{c \in S} d_2(t)g\left(t-cT_c\right) \\ \vdots \\ \sum\limits_{c \in S} d_M(t)g\left(t-cT_c\right)\end{bmatrix}^\mathrm{T}\!\!\!\!,~S=\{0,2,\ldots,C\}
\end{align}
Here the dictionary matrix has been recast into an in-phase version, with pulse shape function included.
Notice that $g(t)$ here and as depicted in \figref{fig:transceiverStruc} is assumed to be a half-sine pulse, which is the pulse shaping function used in IEEE 802.15.4.
This pulse shape has limited support in the time domain, which is not true for e.g. a raised cosine pulse shape.
The equations in this work are defined for the half-sine pulse shape, but are easily changed to apply to other pulse shape functions.

\section{Classic Receiver Structure}
\label{sec:rxStruc}
Before introducing our compressive sensing receiver structure, we first define a classic Nyquist sampling receiver structure.
At the receiver, the received signal is:
\begin{align}
	r_{k}(t) = s_{k}(t) + n(t),
\end{align}
where $n(t)$ is additive white Gaussian noise.

The in-phase and quadrature-phase analog signals are sampled according to the chip rate
using a matched filter to the pulse shape used at the transmitter and an ADC.
Here, we assume a coherent receiver with perfect synchronization, performed prior to data decoding using e.g. a pilot sequence.
The sampling may be represented using a measurement matrix, $\mTheta_1(t)$:
\begin{align}
	y_k^I[\ell] &= \int_{\ell T_c}^{(\ell+1) T_c} \theta_\ell(t) r_{k}^I(t) \mathrm{d}t, \text{ where} \\
    \mTheta_1(t) &= \begin{bmatrix} \theta_0(t) \\ \theta_1(t) \\ \vdots \\ \theta_{C-1}(t) \end{bmatrix}, \begin{array}{l} \theta_i(t) = g(t-iT_c), \\ 0 \leq t < CT_c \end{array}
\end{align}
The measurement matrix is denoted $\mTheta_1$ because it samples every $T_c/1$, i.e. at Nyquist rate.

This means that for every received symbol $2C$ samples must be taken for the in-phase and quadrature-phase signals in total.
These samples then form the received signal vectors, $\vy_{k}^I$ and $\vy_{k}^Q$, 
which are used to demodulate the signal and find an estimate of the transmitted symbol, represented as $\valpha_k$, using a least squares estimator.

Due to the simple design of this signalling scheme and the matched filter, 
it is possible to perform the demodulation process as a least squares estimation with simple purely binary versions of the analog dictionary and measurement matrices, $\mPsi^I(t), \mPsi^Q(t)$ and $\mTheta_1(t)$, respectively. 

Define $\vy_{k} = \vy_{k}^I + j\vy_{k}^Q$ and define $M$ \emph{signal candidates} as $\vs_{m} = \mTheta_1\left(\mPsi^I\valpha_m + j\mPsi^Q\valpha_m\right)$,
where $\mTheta_1 = I$ is now simply the $\mathrm{C\times C}$ identity matrix and $\mPsi^I\in\{\pm 1\}^\mathrm{C\times M}$ and $\mPsi^Q\in\{\pm 1\}^\mathrm{C\times M}$
are the discrete in-phase and quadrature-phase dictionary matrices with each entry signifying either a positive ($1$) or negative ($-1$) pulse in the analog versions of the dictionary matrices.
With these definitions in order the least squares estimate can be found as:
\begin{align}
	\label{eqn:class_rule}\tilde{\valpha}_{k,\text{idx}} = \argmin_{m} \Big(\vy_{k} - \vs_{m}\Big)^\mathrm{H} \Big(\vy_{k} - \vs_{m}\Big)  
\end{align}
where $(\cdot)^\mathrm{H}$ denotes Hermitian transpose,
$\tilde{\alpha}_{k,\text{idx}}$ is the estimate of the index in the $\valpha_k$ vector which is non-zero, i.e. the index corresponding to the symbol that has been transmitted.

\section{Compressive Sensing Receiver Structure}
\label{sec:csStruc}
In hardware compressive sensing sampling structures, such as the Random Demodulator \cite{Tropp2010}, a PRN sequence is mixed with the received signal followed by low-pass filtering.
Due to the presence of a PRN sequence in a spread spectrum transmitter, which spreads the data signal,
a compressive sensing-enabled receiver may merely use a repeated version of its matched filter, subsample the received signal and still demodulate the information.
Before sampling, the matched filter must be modified to contain not only a single chip pulse shape but as many chip pulse shapes as shall be contained per sample.
This received signal vector may then be written as:
\begin{align}
	\vy_k^I[\ell] &= \int_{\ell T_c/\kappa}^{(\ell+1)T_c/\kappa} \theta_\ell(t) r_{k}^I(t) \mathrm{d}t, \text{ where} \\
    \mTheta_{1/\kappa}(t) &= \begin{bmatrix} \theta_0(t) \\ \theta_1(t) \\ \vdots \\ \theta_{L-1}(t) \end{bmatrix}, \begin{array}{l} \theta_j(t) = \sum\limits_{c=j/\kappa}^{(j+1)/\kappa} g(t-cT_c),\\ 0 \leq t \leq CT_c \end{array} 
\end{align}
Here each value of $\ell = 0,1,\ldots,L$ signifies a collection of chips due to the subsampling where $L=C\kappa$ is the number of samples taken per symbol.
$\kappa =\frac{L}{C} \in ~]0,1]$ is the undersampling ratio in the compressive sensing system and signifies the ratio between taken samples and Nyquist samples.
In this work we limit ourselves to scenarios where $1/\kappa$ is an integer number, i.e. only an integer number of Nyquist samples are compressed together into one sample.

To verify that the use of an additional PRN sequence at the receiver is unnecessary, 
we may look at the outcome of the subsampling ADC in \figref{fig:transceiverStruc}.
Assuming a noise-free setting ($n(t)=0$), the outcome becomes:
\begin{align}
	\nonumber y_k^I[\ell] &= \sum_{c=\ell/\kappa}^{(\ell+1)/\kappa} \int_{cT_c}^{(c+1)T_c} r_{k}^I(t)p_{\text{PRN}}(t) \mbox{d}t\\ 
	\nonumber &= \sum_{c=\ell/\kappa}^{(\ell+1)/\kappa} \int_{cT_c}^{(c+1)T_c} \sum_{c'=0}^{C/2-1} b_k(t+c'T_c)c_k(t+c'T_c)\\
	&\qquad \cdot g(t-nT_c)p_{\text{PRN}}(t) \mbox{d}t 
\end{align}
Notice that the up and down-conversions have been assumed perfect and $p_{\text{PRN}}(t)$ is a new PRN sequence, added at the receiver as is done in the Random Demodulator receiver structure \cite{Tropp2010}.
The symbol $c'$ denotes a chip picked out in $d_k(t)$ at the transmitter and used to avoid confusion with $c$, the chips added together into a sample at the receiver.
The special indexing with $T_c$ in connection with $b_k(t)$ and $c_k(t)$ is to pick out the chips in the in-phase path only, similar to what was done in (\ref{eqn:IDic}).
Because everything is multiplicative, it can be seen that $c_k(t+nT_c)$ and $p_{\text{PRN}}(t)$ are synchronized and have the same chip rate, i.e. they may be viewed as a single PRN sequence.
It follows that the multiplication of a PRN sequence at the receiver is unnecessary here.

Because we wish to demodulate a signal, which is equivalent to a classification problem,
it is not necessary for us to reconstruct the full original signal as is done in compressive sensing.
Instead we use the recently introduced concept of compressive signal processing \cite{Davenport2010} to perform classification in the compressed domain.
By classification, we mean to classify which of the signal candidates in the dictionary $\mPsi^I$ and $\mPsi^Q$ has been transmitted.
This does not require reconstruction of the signal itself and may therefore be done with less computational complexity by using compressive signal processing,
rather than classic compressive sensing algorithms, that reconstruct the full signal.

To demodulate the data at the receiver using the two subsampled chip sequences, $\vy_{k}^I$ and $\vy_{k}^Q$,
the classification rule in (\ref{eqn:class_rule}) is used again with $\mTheta_{1/\kappa}\in\{0,1\}^\mathrm{L\times C}$ instead of $\mTheta_{1}\in\{0,1\}^\mathrm{C\times C}$.
In \cite{Davenport2010} a prewhitening matrix, $\mW$, is introduced to counter noise coloring by the measurement matrix.
However, as our proposed measurement matrix, $\mTheta_{1/\kappa}$, has no overlapping rows, the noise remains white in our case.
This prewhitening matrix is therefore not necessary here, but if e.g. a Gaussian or Bernoulli measurement matrix is used instead, it must be included.

\section{Numerical Results}
\label{sec:numExp}
To demonstrate the performance of our proposed receiver structure, we have performed a numerical experiment
in which we compare the Bit Error Rate (BER) of a classical receiver to that of a compressive sensing-enabled receiver.
This is done for a range of Signal-to-Noise-Ratio (SNR) levels.
The system used for this experiment is our MATLAB implementation of the physical layer of the IEEE 802.15.4 $2450\:$MHz OQPSK radio band specification \cite{IEEE802.15.4}.
Each block of four bits is mapped into one of 16 binary chip sequences\footnote{In the published version of this paper we write 32. The correct number is 16.}, according to the mapping in \cite{IEEE802.15.4}.
The chip sequence is then modulated using Offset Quadrature Phase Shift Keying (OQPSK).
This standard has been chosen due to its widespread use, having been deployed already in many applications around the world and because it is a known standard to many scientists and engineers.

The experiment is repeated for a range of SNRs or more specifically energy per bit per noise spectral density (Eb/N0).
The noise is added in a bandwidth corresponding to that of the baseband signal, i.e. $2\:$MHz \cite{IEEE802.15.4}.
Our experiment is conducted by transmitting randomly generated data packets of length $127\times 8=1016$ bits each (the maximum size of an IEEE 802.15.4 data packet).
For each of the two tested methods and for each Eb/N0 level, bits are transmitted until at least $1000$ bits have been received in error.
All MATLAB code developed for this paper is published following the principle of Reproducible Research \cite{Vandewalle2009} and is freely available at \newline {\small\url{http://www.sparsesampling.com/cspdsss2012}}.

To validate the implementation of the compressive sensing framework, we have conducted a numerical experiment in which we added a constant to the transmitted signal, rather than additive white Gaussian noise (AWGN).
The results for both the classical least squares and the compressive sensing implementation follow the expected results as found through mathematical calculations,
thereby indicating that the implementation performs as expected.

\begin{figure}
	\centering
	\includegraphics[width=0.45\textwidth]{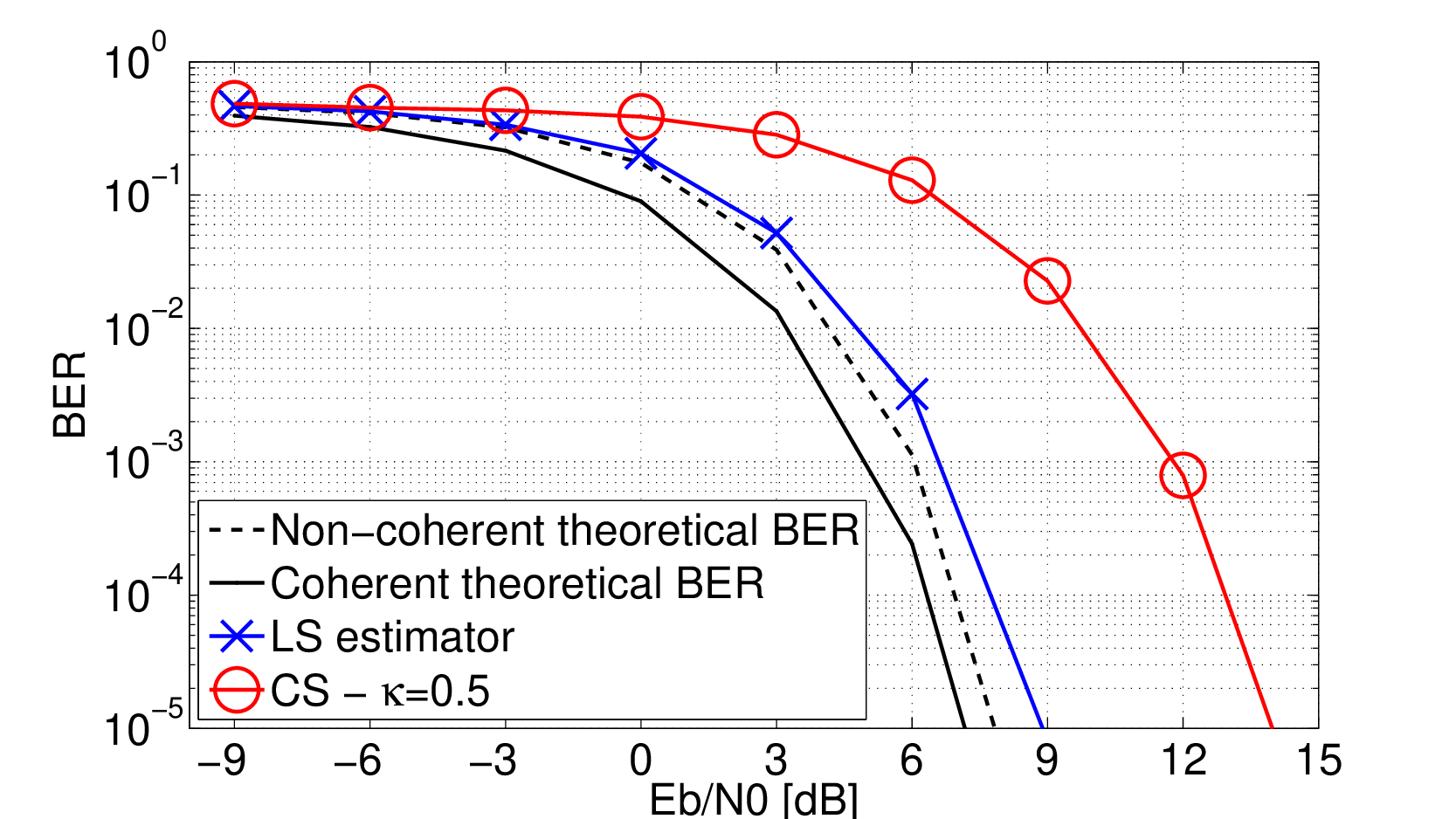}
	\caption{The BER versus Eb/N0 for a classical receiver implementation using least squares compared to that of a compressive sensing enabled receiver with $\kappa=0.5$.
				The full black curve signifies theoretical BER per Eb/N0 for coherent MFSK and the dashed curve is theoretical BER per Eb/N0 for non-coherent MFSK.}
	\label{fig:NumExp_versus_Theo}
\end{figure}

The result of the BER versus Eb/N0 experiment with AWGN is shown in \figref{fig:NumExp_versus_Theo}.
Also shown is the theoretical BER versus Eb/N0 for coherent MFSK \cite{Proakis2008}, numerically evaluated:
\begin{align}
    P_b = \frac{8}{15} \frac{1}{\sqrt{2\pi}} \int_{-\infty}^{\infty}\left[1-\left(1-Q(x)\right)^{15}\right]e^{-\frac{\left(x-\sqrt{8\frac{E_b}{N_0}}\right)^2}{2}}\mathrm{d}x.
\end{align}
We have also included the theoretical curve for non-coherent MFSK, as it is stated in the IEEE 802.15.4 standard \cite{IEEE802.15.4}:
\begin{align}
	P_b = \frac{8}{15} \frac{1}{16} \sum_{m=2}^{16} (-1)^m \left( \begin{array}{c} 16 \\ m \end{array}\right) e^{4 \frac{E_b}{N_0} \left(\frac{1}{m} - 1\right)}
\end{align}
The classical implementation does not follow the theoretical bound exactly because the PRN sequences in \cite{IEEE802.15.4} are not orthogonal and due to the short code lengths.

For $\kappa=0.5$ the compressive sensing receiver performs worse than a classical receiver by $\approx 4$-$5\:$dB,
which is supported by the results on noise folding in \cite{Davenport2012}.

\section{Discussion and Conclusion}
\label{sec:discAndConc}
We have shown that compressive sensing enables subsampling of a DSSS signal.
This has been demonstrated by means of IEEE 802.15.4 2.4GHz OQPSK signals, which we successfully subsampled with half the Nyquist rate.
This subsampling may lead to a decrease in energy consumption or a lowering of the manufacturing price.
The penalty is the expected drop in performance due to noise folding.
This penalty has not been further treated in this work but in \cite{Davenport2012} the authors suggest to incorporate the effect of quantization,
which should favor compressive sensing over a classical receiver as a compressive sensing enabled receiver is able to quantize the signal at a higher resolution,
due to the lower sample rate.

An undersampling of $\kappa=0.5$ is not a large undersampling rate. 
This is mainly due to the effect of noise folding and because the IEEE 802.15.4 standard spread spectrum codes are only $16$ chips long in each channel (I and Q).
For more complex spread spectrum systems with longer chipping sequences (and therefore more potential sparsity) and multiple users and if quantization is included in the signal model, 
we strongly believe there are cases where the sampling rate may be decreased, while still attaining the same or better BER performance than a classical receiver.
This would make compressive signal processing in such systems more attractive.

The main result of this paper is the observation that in a spread spectrum receiver it is possible to use compressive sensing without generating a PRN sequence and mixing it with the received signal.
This is possible because a spread spectrum signal has already been spread by the transmitter.

\bibliographystyle{IEEEbib}
\bibliography{my_bibtex}

\begin{thebibliography}{10}

\bibitem{Candes2006c}
E.~J. Candes, J.K. Romberg, and T.~Tao,
\newblock ``Stable signal recovery from incomplete and inaccurate
  measurements,''
\newblock {\em Communications on Pure and Applied Mathematics}, vol. 59, no. 8,
  pp. 1207--1223, 2006.

\bibitem{Donoho2006}
D.~L. Donoho,
\newblock ``Compressed sensing,''
\newblock {\em IEEE Trans. on Inf. Theory}, vol. 52, no. 4, pp. 1289--1306,
  Apr. 2006.

\bibitem{Davenport2010}
M.~A. Davenport et~al.,
\newblock ``Signal processing with compressive measurements,''
\newblock {\em IEEE J. Sel. Topics Signal Process.}, vol. 4, no. 2, pp.
  445--460, Apr. 2010.

\bibitem{IEEE802.15.4}
{The Institute of Electrical and Electronics Engineers, Inc.},
\newblock {\em {IEEE Std. 802.15.4-2006}}, 2006,
\newblock Available online: \url{http://standards.ieee.com}.

\bibitem{Paredes2007a}
J.~Paredes, G.~Arce, and Z.~Wang,
\newblock ``Ultra-wideband compressed sensing: Channel estimation,''
\newblock {\em IEEE J. Sel. Topics Signal Process.}, vol. 1, no. 3, pp.
  383--395, Oct. 2007.

\bibitem{Zang2009}
P.~Zhang et~al.,
\newblock ``A compressed sensing based ultra-wideband communication system,''
\newblock in {\em Proc. 2009 IEEE International Conf. on Communications},
  Piscataway, NJ, USA, 2009, pp. 4239--4243.

\bibitem{Aggarwal2009}
V.~Aggarwal et~al.,
\newblock ``{{Enhanced CDMA communications using compressed-sensing
  reconstruction methods}},''
\newblock in {\em 47th Annual Allerton Conf. on Communication, Control, and
  Computing}, Sept. 2009, pp. 1211--1215.

\bibitem{Li2011a}
X.~Li et~al.,
\newblock ``{{GPS Signal Acquisition via Compressive Multichannel Sampling}},''
\newblock Available on arXiv: http://arxiv.org/abs/1107.3636v1, 2011.

\bibitem{Pankiewicz2011}
P.~Pankiewicz, T.~Arildsen, and T.~Larsen,
\newblock ``{Sensitivity of the Random Demodulation Framework to Filter
  Tolerances},''
\newblock in {\em 19th European Signal Processing Conf. (EUSIPCO)}, Barcelona,
  Spain, Aug. 2011.

\bibitem{Tropp2010}
J.~A. Tropp et~al.,
\newblock ``Beyond nyquist: Efficient sampling of sparse bandlimited signals,''
\newblock {\em IEEE Trans. on Inf. Theory}, vol. 56, no. 1, pp. 520--544, Jan.
  2010.

\bibitem{Vandewalle2009}
P.~Vandewalle, J.~Kovacevic, and M.~Vetterli,
\newblock ``Reproducible research in signal processing [what, why, and how],''
\newblock {\em IEEE Signal Processing Mag.}, vol. 26, no. 3, pp. 37--47, May
  2009.

\bibitem{Proakis2008}
John~G. Proakis and Masoud Salehi,
\newblock {\em Digital Communications},
\newblock McGraw Hill, 5th edition, 2008.

\bibitem{Davenport2012}
M.~Davenport et~al.,
\newblock ``The pros and cons of compressive sensing for wideband signal
  acquisition: Noise folding vs. dynamic range,''
\newblock {\em to appear in IEEE Trans. Signal Process.}, 2012.

\end{thebibliography}

\end{document}